\documentclass[namedreferences,natbib,etoolbox,final]{solarphysics}
\usepackage[optionalrh,solaromanenum,natbib,linksfromyear]{spr-sola-addons} 	
\usepackage[urlcolor=blue,breaklinks]{hyperref}
\usepackage{graphicx}
\usepackage{rotating}
\usepackage{courier}  				
\usepackage{amssymb}                    	
\usepackage{color}                       	
\usepackage{url}                         	
\ifx \doiurl    \undefined \def \doiurl#1{\href{http://dx.doi.org/#1}{\textsf{DOI}}}\fi
\ifx \adsurl    \undefined \def \adsurl#1{\href{http://adsabs.harvard.edu/abs/#1}{\textsf{ADS}}}\fi
\ifx \arxivurl  \undefined \def \arxivurl#1{\href{http://arxiv.org/abs/#1}{\textsf{arXiv}}}\fi

\newenvironment{acks_M}[1][Acknowledgments]{\footnotesize\paragraph*{#1}}{}
\newcommand{\todash}{\,--\,}

\newcommand{\etal}{{\it et al.}}
\newcommand{\eg}{{\it e.g.}}
\newcommand{\ie}{{\it i.e.}}
\newcommand{\be}{\begin{equation}}
\newcommand{\ee}{\end{equation}}
\newcommand{\beq}{\begin{eqnarray}}
\newcommand{\eeq}{\end{eqnarray}}

\newcommand{\aap}{    {\it Astron. Astrophys.}}

\newcommand{\aapr}{   {\it Astron. Astrophys. Rev.}}

\newcommand{\apj}{    {\it Astrophys. J.}}
\newcommand{\apjl}{   {\it Astrophys. J. Lett.}}

\newcommand{\jgr}{    {\it J. Geophys. Res.}}
\newcommand{\mnras}{  {\it Mon. Not. Roy. Astron. Soc.}}

\newcommand{\solphys}{{\it Solar Phys.}}

\newcommand{\ssr}{    {\it Space Sci. Rev.}}

\newcommand{\RSUN}{R$_{\odot}$}

\newcommand{\kms}{$\mathrm{km~{s}^{-1}}$}
\newcommand{\UOAPhys}{Section of Astrophysics, Astronomy and Mechanics,Department of Physics,University of Athens, Zografos (Athens), GR-15783, Greece}
\newcommand{\UOAInf}{Department of Informatics, University of Athens, Zografos (Athens), GR-15783, Greece}
\newcommand{\UOIPhys}{University of Ioannina, Department of Physics, Section of Astrogeophysics, Ioannina, Greece}
\newcommand{\TEILam}{Department of Electronics, Technological Educational Institute of Lamia, Lamia , GR-35100, Greece}
\newcommand{\NRL}{Space Science Division, Naval Research Laboratory, Washington, DC, USA}

\begin{document}
\begin{article}
\begin{opening}
\title{ CME Expansion as the Driver of Metric Type II Shock Emission as Revealed by Self-Consistent Analysis of High Cadence EUV Images and Radio Spectrograms.}
\author{A.~\surname{Kouloumvakos}$^{1,2}$ \sep
	S.~\surname{Patsourakos}$^{2}$ \sep
	A.~\surname{Hillaris}$^{1}$ \sep
	A.~\surname{Vourlidas}$^{3}$ \sep
	P.~\surname{Preka-Papadema}$^{1}$ \sep
	X.~\surname{Moussas}$^{1}$ \sep 
	C.~\surname{Caroubalos}$^{4}$ \sep
	P.~\surname{Tsitsipis}$^{5}$ \sep
	A.~\surname{Kontogeorgos}$^{5}$
}
\runningauthor{A.~Kouloumvakos \etal}
\runningtitle{The Radio Signature of the 13 June 2010 Impulsive CME}
\institute{$^{1}$ \UOAPhys \\
   email: \href{mailto:athkouloumvakos@gmail.com}{athkouloumvakos@gmail.com} \\
   email: \href{mailto:ahilaris@phys.uoa.gr}{ahilaris@phys.uoa.gr}
				}
\institute{$^{2}$ \UOIPhys~email: \href{mailto:spatsour@cc.uoi.gr}{spatsour@cc.uoi.gr}
			}
\institute{$^{3}$ \NRL}
\institute{$^{4}$ \UOAInf}
\institute{$^{5}$ \TEILam
			}
\begin{abstract} 
On 13 June 2010, an eruptive event occurred near the solar limb. It included a small filament eruption and the onset of a relatively narrow coronal mass ejection (\,CME\,) surrounded by an extreme ultraviolet (\,EUV\,) wave front recorded by the \textit{Solar Dynamics Observatory's} (\,SDO\,) \textit{Atmospheric Imaging Assembly} (\,AIA\,) at high cadence. The ejection was accompanied by a GOES M1.0 soft X-ray flare and a Type-II  radio burst; high-resolution dynamic spectra of the latter were obtained by the \textit{Appareil de Routine pour le Traitement et l'Enregistrement Magnetique de l'Information Spectral} (ARTEMIS IV) radio spectrograph. The combined observations enabled a study of the evolution of the ejecta and the EUV wavefront and its relationship with the coronal shock manifesting itself as metric Type-II burst. By introducing a novel technique, which deduces a proxy of the EUV compression ratio from AIA imaging data and compares it with the compression ratio deduced from the band-split of the Type-II metric radio burst, we are able to infer the potential source locations of the radio emission of the shock on that AIA images. Our results indicate that the expansion of the CME ejecta is the source for both EUV and radio shock emissions. Early in the CME expansion phase, the Type-II burst seems to originate in the sheath region between the EUV bubble and the EUV shock front in both radial and lateral directions. This suggests that both the nose and the flanks of the expanding bubble could have driven the shock.
\end{abstract}
\keywords{Coronal Mass Ejections: Low Coronal Signatures; 
		Corona: Radio Emission;
		Radio Bursts: Meter-Wavelengths and Longer (m, dkm, hm, km);
		Radio Bursts:  Type II;
}
\end{opening}

\section{Introduction}\label{intro}

Observations of frequency drifting radio sources have provided indirect evidence for the existence of shocks in the low corona for over sixty years \citep{Wild1950}. The radio emission from these, so-called Type-II, sources is thought to originate at the local plasma frequency and/or its harmonics via plasma waves excited by electrons accelerated at a shock. As the coronal electron density, and the local plasma frequency, drop with height, the Type-II emission drifts to lower frequencies  \citep[$\frac{1}{f} \frac{\mathrm{d}f}{\mathrm{d}t} \approx 0.001$\todash{}$ 0.01 \rm \; Hz$; \eg {~Table~A.1 of}][]{Caroubalos04}. The speed of the exciter can then be estimated from the frequency drift rate of the burst, if the coronal density gradient is known or can be assumed. Type-II sources emit from high frequencies ($\approx 800$ MHz) \citep{2011SPD....42.1307W,2012ApJ...746..152M} deep in the corona to the kHz range at 1 AU. Their emission is frequently intermittent, especially at the lower frequencies, complicating the association of a particular Type II across multiple frequencies and instruments. 


Type-II spectra sometimes show a split in two of a given harmonic (or fundamental or, sometimes, both) lanes \citep{1967PASAu...1...47M}. These so-called split-band Type IIs have the same drift rate and overall morphology, but they are separated by a small frequency offset of $\approx f/8$\todash{}$f/4$, which increases with frequency \citep{1974IAUS...57..389S}. Their origin, within a structured ambient environment, may be similar to the origin of the multiple-lane Type IIs discussed above. An alternative interpretation attributes the emission to electrons accelerated in the upstream and downstream region of the shock. This idea, first proposed by \citet{1974IAUS...57..389S}, has gained popularity because it allows the inference of physical quantities such as upstream magnetic field and shock compression ratio, which are in general agreement with theoretical expectactions for those coronal regions \citep[\eg][]{1974IAUS...57..389S,Vrsnak01a,Vrsnak02,Vrsnak04}. Apparent observational support of the \citet{1974IAUS...57..389S} band-splitting interpretation has recently been provided by \citet{Zimovets2012} based on combined \textit{Solar Dynamics Observatory's} \citep[SDO:][]{Pesnell2012} \textit{Atmospheric Imaging Assembly} \citep[AIA:][]{Lemen2012} and \textit{Nan\c cay Radioheliograph} recordings of a limb event on 3 November 2010. Alternatively, \cite{Kumar2013} showed evidence of blast wave for the same event, with their speed matching the speed of Type-II source. Other theoretical and observations considerations, however, \citep[see][]{1992ApJ...399L.167T,Grechnev2011} and recent simulations \citep{2008A&A...478L..15S} do not support this interpretation. Without a widely accepted mechanism for the interpretation of split-band Type IIs to fall back on, we chose to adopt the \citet{1974IAUS...57..389S} interpretation for part of our analysis while acknowledging its limitations. It is our opinion that the full benefit of Type-II observations can only come about through detailed modelling of both the radio emission and the driver \citep[][~and references therein]{2012JGRA..11703104H}.

Metric Type IIs occur in the low corona (frequencies above 300 MHz, corresponding to heights $<$\,0.1\todash{}0.3 \RSUN; see for example the review by~\citet{Vrsnak08}) and normally last for only a few minutes \citep{Pick08}. A serious obstacle in our understanding of metric Type IIs is the uncertainty in the nature of their drivers. It has long been established that the large scale, complex magnetic field plasma structures ejected from the Sun, known as coronal mass ejections (CMEs), are the drivers of the shocks behind Type IIs at decimetric or longer wavelengths \citep{1987JGR....92.9869C,1999JGR...10416979R}. However, the nature of the Type-II driver at metric wavelengths remains unclear. The metric Type-II bursts are variously thought to be either CME-driven shock signatures \citep{Kahler84,Maia00,Classen} or flare blast shock  emissions \citep{Vrsnak01,2001JGR...10625301L}. The problem arises from the lack of imaging observations of metric Type-II sources and from discrepancies among the speeds of CMEs, and metric and decimetric bursts \citep[\eg][]{2000JGR...10518225L,2001JGR...10625279R}.

Radio imaging can supply important and valuable information on metric Type IIs \citep[\eg][]{Hudson07},~although with a significant limitation. Typically, the spatial resolution of such observations is of the order of several tens of arcsec, which means that only the crude spatial characteristics of the exciter-shock system can be studied. Extreme ultraviolet (\,EUV\,) and soft X-ray (\,SXR\,) imaging observations of events associated with metric Type IIs are then particularly well suited for that task since they supply a superior spatial resolution ($\approx$ few arcsec). Moreover, given the  short duration of metric Type IIs, ultra-high cadence  EUV or SXR observations are required if we want to trace in time the evolution of these phenomena.

Thus, in order to  clarify the association between radio emission and erupting features in other wavelengths, we need to relate the onset and duration of metric Type-II emission (which is usually recorded with high cadence and spectral resolution) with the time history and positional information of flares and CMEs in the low corona (where high spatial and temporal resolution have been rare). Thankfully, ultra-high  cadence, arcsecond-level EUV full-disk imaging has become routinely available with the operation of the AIA onboard the SDO. The 12-second cadence of AIA images is sufficient to temporally resolve the entire life time of short-lived phenomena such as metric Type IIs and opens a new and powerful window for the study of the sources of low coronal shocks. For the breadth of  wave and shock  phenomena and their drivers as observed in the EUV, the reader could consult the recent review by \citet{Patsourakos2012}.

 An eruptive GOES class M1.0 flare, took place on 13 June 2010 and provided us with an optimal combination of high-cadence EUV imaging and high-resolution radio spectra for investigating the nature of the drivers of metric Type IIs. The mass ejection observed in the low corona by AIA had the form of an EUV bubble, which eventually evolved into  a narrow CME observed with coronagraphs \citep[][]{Patsourakos2010}. The expanding EUV bubble launched a propagating intensity disturbance  around it (\ie~an EUV wave), which was probably a shock. Finally, a metric Type-II radio burst exhibiting a fundamental--harmonic (F--H) structure took place during the event \citep[see, for example,~][]{Kozarev2011,Ma2011,Gopalswamy2012}. 

In this article, we take advantage of  high resolution spectra from the \textit{Appareil de Routine pour le Traitement et l' Enregistrement Magnetique de l' Information Spectral} \citep[ARTEMIS IV:~][]{Caroubalos01,Kontogeorgos, Kontogeorgos06, Kontogeorgos08} multichannel radio spectrograph. As the ARTEMIS IV range extends to higher frequencies than the \textit{Radio Solar Telescope Network} ~\citep[RSTN:~][]{Guidice81} recordings, it is possible to observe the start of the Type II that appear above 180~MHz. 


We study here the characteristics of shock formation and propagation of the Type-II burst. We examine the relationship between this burst  and its exciter (the EUV bubble, as we will see later) using detailed kinematic profiles for the EUV structures. We also introduce a new technique to relate radio to EUV structures by comparing the compression ratio calculated in the EUV to the compression ratio corresponding to the band split of the Type-II radio spectrum based on the \citet{1974IAUS...57..389S} assumption.

The article is structured as follows. In Section \ref{Event} we describe the data sets and instruments used in our study. We then establish the relationship between the Type-II and the EUV emissions using the kinematic information (Section~\ref{HT}) and the corresponding compression ratios (Section~\ref{CRatio}). We discuss the results and conclusions in Section~\ref{Disc} 

\begin{figure}\centering
\centerline{	\includegraphics[width=0.45 \textwidth]{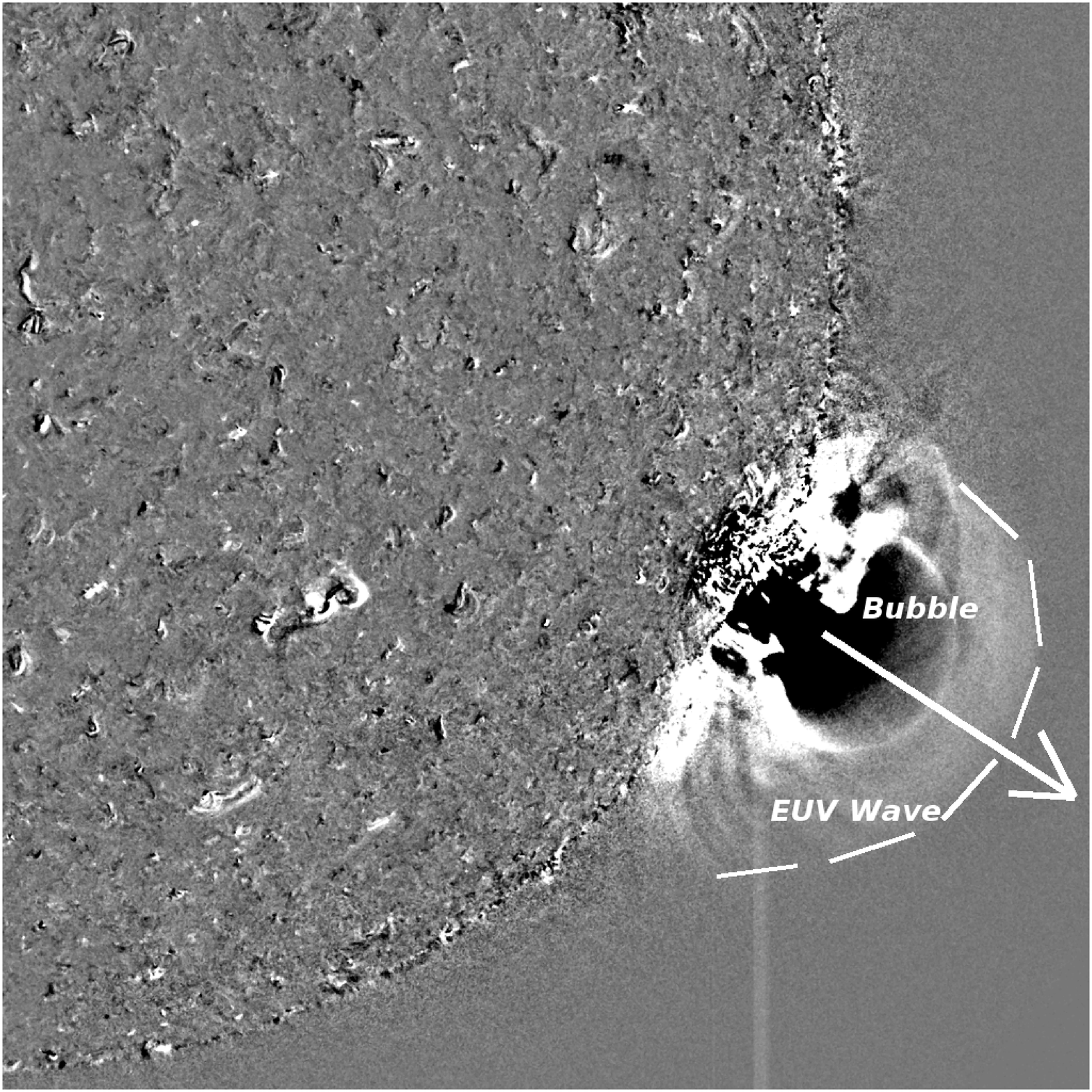}}
  \caption{AIA base-difference image at 211\,\AA, during the bubble initial expansion at 05:40 UT (reference image at 05:30 UT); the thick white arrow points in the direction of the radial expansion. The regions of the bubble and EUV wave are indicated; the end of EUV wave is marked with a dashed line.}
  \label{SynthAIA}
\end{figure}

\begin{figure}\centering
\centerline{\includegraphics[width=0.85 \textwidth]{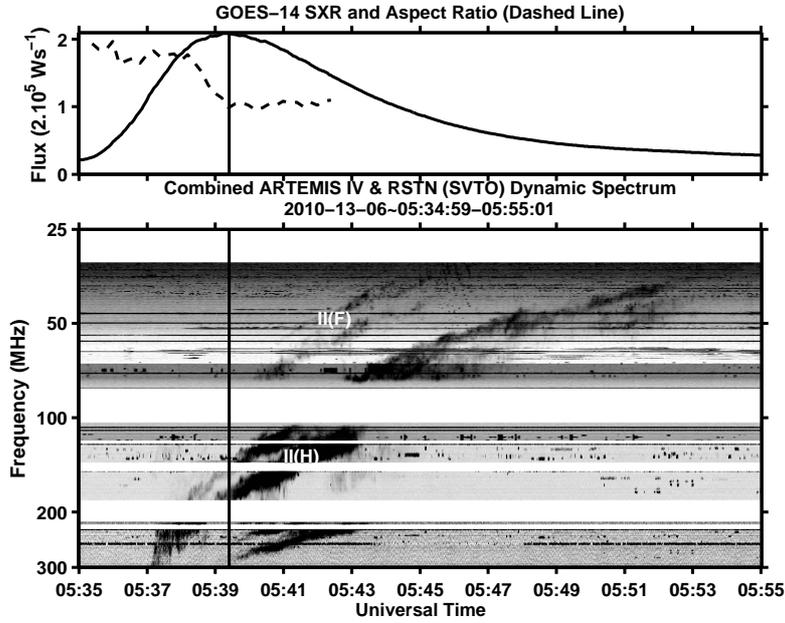}}
   \caption{ Top: GOES--14 SXR flux in the 0.5\todash{}4\,\AA\, chanel and EUV bubble aspect ratio (dashed line).  The vertical line at 05:39 UT marks at the transition from aspect ratio $\approx$\,two to $\approx$\,one marking the start of the EUV bubble lateral expansion phase (see Table \ref{table:1}). Bottom: Combined dynamic spectra of ARTEMIS IV and RSTN (San Vito) in the 25\todash{}300 MHz range; the fundamental and the harmonic of the Type-II burst are annotated with II(F)  and II(H) respectively. The Type-II bands below the II(H) in the range 275\todash{}300 MHz were found to be an  artefact due to the non-linear response of the ARTEMIS IV pre-amplifier which introduced an image of II(H) displaced by 100 MHz. (The horizontal black stripes are interference from terrestrial emitters; the white stripes are disturbed frequency ranges that have been suppressed.)}
  \label{MultiFlareWave}
\end{figure}

\section{Observations and Data Analysis} \label{Event}


For the analysis of the 13 June 2010 event, we use dynamic radio spectra from the ARTEMIS IV radio spectrograph based at Thermopylae, Greece (www.cc.uoa.gr /$\sim$artemis/), in the range 100\todash{}300 MHz with time resolution of 0.1 seconds. These are supplemented with spectra from the San Vito Solar Observatory of the RSTN \href{http://www.ngdc.noaa.gov}{(www.ngdc.noaa.gov)}, in the range 25\todash{}100 MHz. We use full-disk EUV images from the AIA imaging suite centered at 171, 193, 211, and 335 {\AA}\,.
AIA takes images with a 1.5-arcsecond spatial resolution and a 12-second temporal cadence \citep{Lemen2012}.

\subsection{Event Overview}\label{Overview}

We analyse an event which took place on 13 June 2010 in NOAA active region 11079 (S25$^\circ$~W84$^\circ$). The temporal evolution of our event and other associated activities recorded in various spectral and spatial domains is summarized in Table~\ref{table:1}. From the analysis of the high-cadence AIA images, \citet{Patsourakos2010} provided a detailed examination of the formation and evolution of the EUV bubble, which reached a maximum speed of $\approx$ 400 \kms. The bubble was formed at 05:35 UT from a set of slowly rising loops, started to expand outwards at 05:36 UT, and underwent a short-lived strong lateral over-expansion starting at 05:38 UT. The bubble  exited the AIA field of view at 05:45 UT and evolved into a narrow CME~(angular width~$\approx$\,33$^\circ$) that was recorded by the SOHO/LASCO coronagraph with an average velocity of $V_\mathrm{CME}\approx$ 320 \kms at a position angle of $250^\circ$~\citep{Patsourakos2010,Gopalswamy2012}. The mass eruption was accompanied by an M1.0 flare between 05:30\todash{}06:40 UT with peak flux at 05:39 UT and a small filament eruption that started at 05:32 UT~\citep{Patsourakos2010}.

The bubble expansion is best observed in 171\,\AA~and~193\,\AA~AIA channels, which correspond to coronal plasma at temperatures of 0.8 and 1.25 MK respectively. A darker expanding area surrounding the bubble is also observed in 171\,\AA~during the expansion phase. This area appears as a bright propagating wave disturbance, \ie~an EUV wave, in the hotter  211\,\AA~and~335\,\AA~AIA channels, with peak responses at 1.6 and 1.7~MK, and it is presumably driven by the bubble expansion (Figure~\ref{SynthAIA}; see the two movies of running-difference images, in different AIA channels during the bubble expansion, in the Electronic Supplementary Materials). 

In the composite dynamic spectra of ARTEMIS IV and RSTN (San Vito) the start of the Type-II radio burst (Figure \ref{MultiFlareWave}, at $\approx$\,05:37~UT) at 300 MHz (first harmonic) and 150 MHz (fundamental) coincides with the onset of the EUV wave which appears one minute after the bubble formation, at 05:37 UT. The fundamental and harmonic radio emission of the Type-II burst are labelled II(F) and II(H) respectively; they were also recorded by the Hiraiso radio spectrograph and reported by \citet{Gopalswamy2012}. The Type-II burst drifts toward lower frequencies ending at 20 MHz (harmonic)\todash{}05:53 UT and its overall duration was $\approx$\,16 minutes. We can estimate the duration of the burst reliably since there was no hectometric extension on the \textit{Wind\//Waves} \citep{Bougeret1995} spectra. Note that no Type-III radio bursts were observed during this event.

Using images in the  171\,\AA\, channel of AIA, \citet{Patsourakos2010} have calculated the aspect ratio of the bubble. They fitted circles to the bubble outline during the time interval the bubble was entirely in the AIA field of view. The ratio of the best-fit bubble height to radius was  taken as the aspect ratio and can be considered as measure of the bubble expansion in the radial and lateral directions. Within the 05:38\todash{}05:39 UT interval the aspect ratio drops from $\approx$\,two to $\approx$\,one (see vertical line on Figure \ref{MultiFlareWave}). This indicates that the EUV bubble enters a strong lateral expansion phase during that interval. We note that the Type-II radio burst also starts around the bubble lateral over-expansion, which suggests that this phenomenon could play an important role into driving the shock. In the next sections we look for spatial and temporal connections between the low coronal shock observed in radio and the structures and evolutions observed by AIA that were described above.

\begin{table}
\centering \caption{\textbf{Overview of the 13 June 2010 event and associated activity.}}
\label{table:1}
\begin{tabular}{lllp{4.95cm}}

\hline
\textbf{Event }       	& \textbf{UT}		& \textbf{Characteristics } 	& \textbf{Remarks }		\\
\hline
CME Onset 		& 05:15			& 				&SOHO/LASCO, PA:~250$^\circ$	\\
 (estimated)		&  			& 				&  	 Width:~33$^\circ$		\\
SXR Start    		& 05:30 		& AR 11079: 			& GOES 14 			\\
  			&  			& S25$^\circ$~W84$^\circ$		&  			\\
Filament 		& 05:32			&				&\citep{Ma2011}	\\
Activation		&  			& 				&  			\\
\hline
Bubble Formation	& 05:35 		& 				& SDO/AIA 			\\	
Bubble Expansion 	& 05:36 		& 225  \kms			& Self--Similar  Radial Expansion \\
 Starts			&  			& 				&  			\\
EUV Shock front		& 05:37 		& 740 \kms			& \citep{Kozarev2011}		\\
appearance		&                      	&                      		&  193/211~\AA~SDO/AIA 	\\
Type II  Start		& 05:37 		& 700 \kms			& ARTEMIS IV and RSTN		\\
			&                      	& 150/300 MHz         		& Fundamental--Harmonic	(ART.~IV)	\\
\hline
Non--linear bubble 	& 05:38 		& 				& Lateral  Expansion 		\\
expansion starts	&                      	& 400 \kms  			& Front \citep{Patsourakos2010} \\
			&                      	& 300 \kms   			& Bubble Flank at $\approx30^\circ$ 	\\
SXR Peak		& 05:39			& M1.0				& GOES 14  			\\
Type II		& 05:40         	& $\approx $450\todash{}700 \kms   			& Down -- Up lane 	\\
\hline
EUV Wavefront		& 05:42 		& ------			& Exits SDO FOV 		\\
Bubble Expansion	& 05:45 		& 300 \kms			& Exits SDO FOV			\\
Type II  End		& 05:53 		& 20 MHz 			& 				\\
\hline
CME C2 			& 06:06			& 320 \kms			& SOHO/LASCO \\
 Appearance		&  			& (Linear Speed)		&  			\\
 SXR End       		& 06:40 		&				& GOES 14  			\\
\hline
\end{tabular}
\end{table}

\section{ Height--Time and Velocity Measurements from the Radio Spectra and SDO/AIA Images}\label{HT}

In this section we associate the EUV bubble and wave kinematics with the kinematics of the radio shock corresponding to the Type II. Our aim is to connect the sources of the radio shocks with features observed in the EUV images.

\subsection{Coronal Density--Height Model Selection} \label{Density}
As plasma emission depends on electron density, which in turn may be converted to coronal height (or conversely heliocentric distance) using density models, we can estimate the radio-source heights and speeds from dynamic spectra. The establishment of a correspondence between frequency of observation--coronal height and frequency drift rate--radial speed is affected by ambiguities introduced by the variation of the ambient medium properties. These may be the result of the burst-exciter propagation within undisturbed plasma, over-dense or under-dense structures or post-CME flows. Several phenomenological models have been introduced
to describe the variation of electron density [\,$n$\,] with the heliocentric distance [\,$R$\,].  

The $n(R)$ functions are exponential \citep{Newkirk}, or more frequently, finite sums of
power-law terms in (\,\RSUN/\textit{R}\,) \citep[see~][]{Allen1947,Saito1977,Leblanc1998,Vrsnak04}.  
\citet{Mann99}  have adopted a different approach based on the \citet{Parker1958} 
equations for the solar wind and the corona. Finally, \citet{Cairns2009}, making the assumption that coronal
plasma flows into the solar wind along conical magnetic funnels, introduced a density--height model:

\be
\centering
{n_\mathrm{e}}\left( R \right) = C {\left( {R - \mathrm{R_ \odot }} \right)^{ - 2}} 
\ee

\noindent where the constant [\,C\,] accounts for differences in the slowly varying outflow speed of the plasma. In general this model does not contradict the existing empirical models \citep[see\,][\,their Figure 4]{Cairns2009}, yet it provides a physically justified density--height relationship, and it is used throughout this article. Note that the \citet{Cairns2009}  density model was found in a good agreement with the drift-rates of several observed metric Type-III radio bursts, occurring at similar frequency ranges (thus heights) as the 13 June 2010 event.

Using this density model, we computed the height--time plots (Figure~\ref{HeightBubble}, grey colour shaded area). In Table~\ref{table:1} we supply the  Type-II bursts speeds from linear fits to the deduced 
height--time pairs.

\subsection{Comparison of Shock Propagation with Type-II Dynamic Spectrum} \label{TypeII}
Given that our event took place at the limb, we can assume that projection effects should have a rather small impact on the determination of the heights associated with several  observed features of the event such as the EUV bubble and wave. 

\begin{figure}\centering
\includegraphics[scale=0.65]{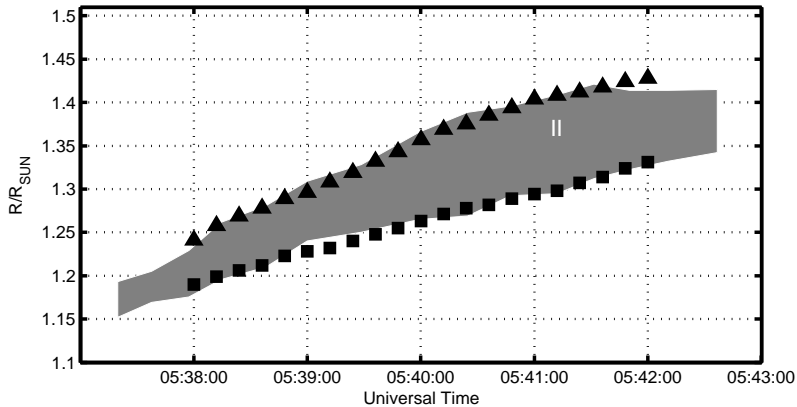}
\caption{Comparison of dynamic spectrum for the Type-II burst harmonic band (in gray) with the height--time profiles of the bubble and EUV wave (see discussion in text). The bubble and EUV wave-front trajectories in 211\,\AA\, AIA channel are shown as squares and  triangles respectively. The frequency--time spectrum has been converted to height--time using the density model of \citet{Cairns2009}. }
 \label{HeightBubble}
\end{figure}

From the SDO/AIA images in 211\,\AA~we estimated the height of both the bubble and the EUV wave. 
Different features, depending on temperature, are best observed in different channels. The bubble appears best in 193\,\AA~and 211\,\AA, while the EUV wave is most clearly seen in 211\,\AA~and in 335\,\AA; in the latter we have enhanced S/N using smoothing and median filtering. The bubble and the wave start forming at about 1.2~\RSUN~ and 1.25~\RSUN~ respectively and continue an almost self-similar expansion until 05:45 UT; they both leave the SDO/AIA field of view at about this time. The results of the comparison of dynamic spectra of the Type-II burst with the height--time profiles of the bubble and wave are depicted in Figure \ref{HeightBubble}. We observe that the Type-II formation is connected with the region between the bubble (EUV wave driver) and the EUV shock front along the radial direction. However the selection of the model constant [\,C\,] is bound to introduce some uncertainty as it depends on the unspecified plasma outflow speed \citep[see discussion by][]{Cairns2009}.

\section{Identification of the Shock Formation from the Calculation of the Compression Ratio}\label{CRatio}

In this section we estimate and compare the compression ratio from both the radio and EUV data, exploiting the band splitting of the Type II and by introducing a new method to compute a proxy for the compression ratio from the EUV images. Our aim is to identify the potential source regions of the Type II on the EUV images, and to relate them with various observed features like the EUV bubble and wave.

\subsection{Estimation of the Compression Ratio from the Band Split of the Type II}
We start with the estimation of the compression ratio from the band splitting of the Type-II lanes. If $f_\mathrm{d}$ corresponds to the lower frequency branch and $f_\mathrm{u}$ to the higher frequency branch, then  the compression ratio is $X=(f_\mathrm{u}/f_\mathrm{d})^2$ since $f_\mathrm{p} \propto {n_\mathrm{(R)}}^{1/2}$.

\begin{figure}\centering
\centerline{ \includegraphics[width=0.60 \textwidth]{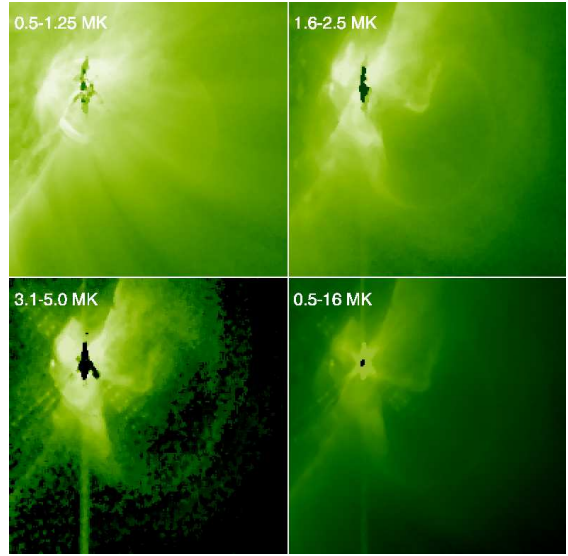}}
   \caption{DEM images resulting from the total DEM within the indicated temperature ranges. Calculations refer to 05:40 UT.}
         \label{Demprev}
   \end{figure}

The compression ratio  for the Type II was calculated in the interval (05:38\todash{}05:41 UT) that coincides with the period of strong lateral expansion of the EUV bubble; before that time the band splitting was not clearly discernible and around ~05:42 UT the EUV wave exited the field of view of AIA. Our compression ratios as derived by the Type-II band splitting was computed in the range $X_\mathrm{{radio}}\approx1.4$\todash$1.5$ and these values are consistent with the results of \citet{Gopalswamy2012}, which report X in the 1.42\todash1.60 range.

\begin{figure}\centering
\centerline{ \includegraphics[width=1.2 \textwidth]{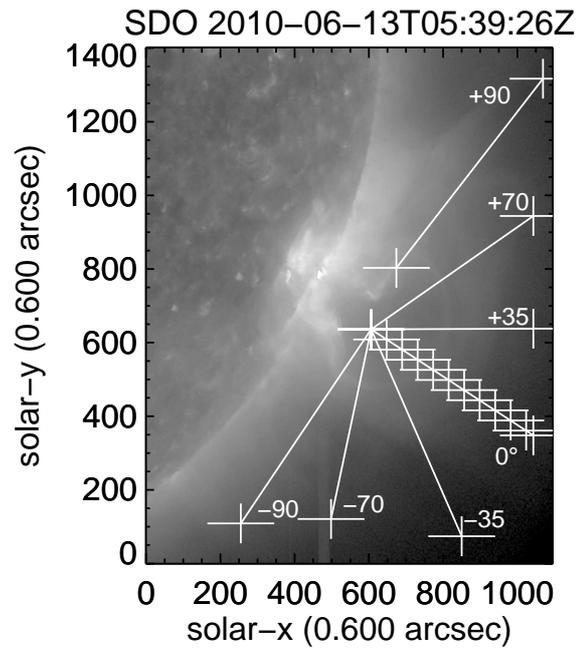}}
   \caption{AIA 211\,\AA~images during the bubble and EUV wave expansion at 05:39:26 UT. The lines label the different directions where the compression ratio profiles were computed. We define the direction of the bubble radial expansion as the zero angle [$\theta=0^\circ$] profile; directions on the North of the radial direction correspond to positive angles. For $\theta=+90^\circ$ we intentionally chose to start the computation further from the standard point, to avoid the rise component of the loop for that direction. For $\theta=0^\circ$ we also mark with crosses every 50 pixels.}
         \label{PrevImgCompRatioDirect}
   \end{figure}

From the Type-II shock compression ratio [\,X\,] we calculated an estimate for the Alfv\'{e}n Mach Number [\,M$_\mathrm{A}$\,] of the ambient plasma \citep[see:~][]{Vrsnak02}.
\be
\centering
 M_\mathrm{A}  = \sqrt {{{X \left( {X  + 5} \right)}}/{{2\left( {4 - X } \right)}}} 
\ee
\noindent assuming perpendicular shock propagation and vanishingly small plasma beta. The Mach number is thus estimated in the range 1.3\todash{}1.5 and for a shock speed $\approx$ 700\,\kms corresponds to an ambient magnetic field 1.7\todash{}1.9 Gauss.

\subsection{Estimation of the Compression Ratio from EUV AIA Images}\label{EUVcompDisc}
We now proceed with the calculation of a compression ratio proxy from the EUV images of AIA. This is particularly useful, since it allows to search for candidate sites from which the Type II of our event could originate. We define $X_\mathrm{{radio}}$ and $X_\mathrm{{EUV}}$ as the compression ratios from radio and EUV respectively.

The EUV intensities that AIA records are essentially the line-of-sight integral of $n^{2}$, with  $n$ the electron density, times the temperature-response function of the channel under consideration. If we omit for the moment the EUV-intensity dependence on temperature, taking the intensity profile in a direction where the bubble expands during the event and calculating the square root of the ratio between the intensity profile [\,$n_{2i}$\,] and the intensity of the upstream region [$n_{1}$] at the pre-event image, we calculated a proxy for the compression ratio in the EUV, [\,$ X_\mathrm{{EUV}}$\,]. Calculating $ X_\mathrm{{EUV}}$ in the region between the bubble and the inner boundary of the EUV wave upstream region, gives a proxy of the compression profile in the wave sheath region (\ie~ down-stream from the wave). We therefore have that both $X_\mathrm{{radio}}$ and $ X_\mathrm{{EUV}}$ are proxies of the compression ratio in regions upstream and downstream of the shock wave.

During the event plasma heating around and between the expanding bubble and the EUV shock was observed \citep[\eg][]{Kozarev2011,Ma2011} and  plasma was brought from the temperature of peak response of the 171\,\AA\,channel to the characteristic temperatures of the hotter 193, 211, and 335\,\AA\,channels. Therefore the $X_\mathrm{{EUV}}$ obtained for the latter channels should be viewed as an {\it upper} limit for the compression ratio, a sort of {\it isothermal} compression ratio.

\begin{figure}\centering
\centerline{  \includegraphics[width=0.90 \textwidth]{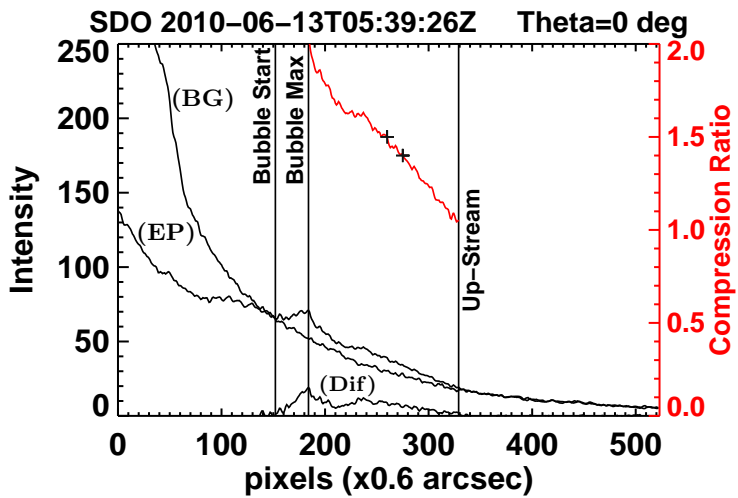} }
       \vspace{-0.320\textwidth}   
     \centerline{\small \bf     
         \hspace{0.218\textwidth}  \color{black}{(EP)}
      \hfill}
       \vspace{-0.180\textwidth}   
      \centerline{\small \bf     
         \hspace{0.255\textwidth}  \color{black}{(BG)}
      \hfill}
       \vspace{0.275\textwidth}   
      \centerline{\footnotesize \bf     
         \hspace{0.415\textwidth}  \color{black}{(Dif)}
      \hfill}
     \vspace{0.110\textwidth}
   \caption{Measurement of the EUV-intensity profile and the corresponding compression ratio along the $\theta=0^\circ$ direction for the 211\,\AA, at 05:39:26 UT. A pre-event background intensity profile at 05:30 UT is displayed with the ``BG" label, the intensity profile during a snapshot (05:39:26) of the event is displayed with the ``EP" label and its difference with the latter (=EP-Dif) presented by the ``Dif" label. The vertical lines from left to right correspond to the start of the bubble, the bubble maximum and the inner boundary of the sheath region (see also the discussion of Section~\ref{EUVcompDisc}). The calculated EUV compression ratio [\,$X_\mathrm{{EUV}}$\,] for the region between the bubble maximum and the wave upstream region is displayed with the red line. The black crosses mark the measured interval of $X_\mathrm{{radio}}$ values during the observed Type II. The $x$-axes is in pixels along the corresponding path (see Figure~\ref{PrevImgCompRatioDirect}).}
         \label{PrevProfileDirect}
   \end{figure}

 We also performed a differential emission measure (DEM) analysis of the entire field covering both the EUV bubble and wave. For this task, we used the AIA images in all six coronal channels (94, 131, 171, 193, 211, and 335\,\AA) taken around 05:40. The method of \citet{2013ApJ...771....2P} was used in the DEM calculation. From Figure~\ref{Demprev}, where  we display the total DEM for selected temperature ranges, we note that the bubble and wave are better traced in the DEM ``image" for 1.6\todash{}2.5 MK. This essentially means that the temperature of the bubble and more importantly of the wave lie within that region. We considered full cadence and resolution 193, 211, and 335~\AA~AIA images, but the presentation of the final results is focused on 211~\AA~where both the EUV bubble and wave are best observed. Our DEM analysis confirms the formation of EUV bubble and wave, within the temperature range of the 211~\AA~channel.

We now proceed to the details of our calculations of $X_\mathrm{{EUV}}$ from the AIA images. For our calculations, in the EUV, we used 211~\AA~ images from 05:38:02 to 05:40:26 UT. We defined different directions (Figure~\ref{PrevImgCompRatioDirect}) in both sides of the radial bubble expansion (\eg~$\pm35^\circ, \pm70^\circ, \pm90^\circ$), where we measured the intensity profiles. Firstly, we smoothed the images with a box-car window of three full-resolution pixels wide. This procedure significantly enhanced the S/N ratio permitting observation of the bubble and the shock front at greater distances. To further improve the S/N of the profiles we took the average intensity for 20 pixels (\ie~ten pixels at each side) across the profile direction. For each direction we label in Figure~\ref{PrevImgCompRatioDirect} we measured the intensity profile from the pre-event image (05:30:00 UT) and the corresponding intensity profile during several snapshots during the event. For the image at 05:39:26 UT and the direction $\theta=0^\circ$, the resulting intensity profiles of the background and the event are labelled as ``BG" and ``EP" respectively in Figure~\ref{PrevProfileDirect}. We then calculated the difference between the background profile and the event profile (containing both the bubble and the wave), labelled as ``Dif" in Figure~\ref{PrevProfileDirect}.

From the difference curve (Figure~\ref{PrevProfileDirect}), we are able to identify different features that are present during the bubble expansion. It is important to identify the region of the bubble and the region of the EUV wave. At the difference curve we observe a maximum which corresponds to the bubble and further away a secondary weaker peak which corresponds to the EUV wave. The point where the event intensity initially overcomes the background (difference curve starts to rise from zero), is the bubble start. The point where the event intensity reaches the background for the second time and both are almost equal (difference curve drops to zero), is the wave end. The region where the wave ends can be defined as the inner boundary of the upstream region of the shock wave, and this region is just perturbed by the shock. The intensity of the upstream point will be used in the calculation of the $ X_\mathrm{EUV}$. 

To automate the selection of the different regions of bubble start and the wave end we used the Poisson--CUSUM method. Cumulative sum (CUSUM) quality-control schemes were proposed by \citet{Page1954} and are widely used in other fields, such as solar energetic particle onset time determination \citep{2005A&A...442..673H}. A CUSUM control scheme cumulates the difference between an observed and a reference value. If this cumulation equals or exceeds a decision interval value, then an out-of-control signal is given at the exact moment when the process transition to the observed value has happened. In our case, the transition is the bubble start and the wave end causing intensities to rise above or below the pre-determined background.

To connect $X_\mathrm{radio}$, which is defined as  $X=(f_\mathrm{u}/f_\mathrm{d})^2 \approx n_\mathrm{d}/n_\mathrm{u}$ (where $n_\mathrm{d}$ and $n_\mathrm{u}$ are the approximate upstream and downstream densities and the $f_\mathrm{u}$ and $f_\mathrm{d}$ are the measured frequencies of the up and down lane of the Type II), with the $ X_\mathrm{EUV}$ we need proxies for the densities of the upstream and downstream region. The latter, [$n_\mathrm{d}$ and $n_\mathrm{u}$] are the approximate densities of the upstream and downstream region. Similar methods to deduce a shock wave compression ratio have been applied by \citet{2009ApJ...693..267O}, where from mass profiles of calibrated LASCO images that they derived some estimates of the density profile across white-light shock fronts.

From the discussion above we have that $ X_\mathrm{EUV}\approx  \sqrt{I_\mathrm{d}~/~I_\mathrm{u}}$ if we assume that the line of sight dependence of the EUV intensity is the same for the up-stream and down-stream regions when taking the corresponding intensity ratios. This is a zero-order approach, since the observations did not allow for a direct determination of the LOS extent. However, given the small size of the ``search" region, \ie~the sheath region between the bubble and the EUV shock, this is a reasonable assumption. The intensity of the upstream region can be easily determined by the method we previously introduced and corresponds to the intensity of the wave end (\ie~the inner boundary of the upstream region; see also Figure~\ref{PrevProfileDirect}). The outer limit of this range gives also a proxy for the upstream quantities.  

Then, the intensity of the downstream region is given by $I_\mathrm{d}(i)$, where ``i" corresponds to the range between the bubble maximum and the wave, and that of the inner boundary of the upstream region is given by $I_\mathrm{u}$. Thus, we have a $ X_\mathrm{EUV}$ profile for the downstream region which is $ X_\mathrm{EUV}=n_{2}(i)/n_1 \approx \sqrt{I_{d}(i)/I_{u}}$. To compare the $ X_\mathrm{EUV}$ values with the $X_\mathrm{radio}$ from the Type-II band splitting we mark at the $ X_\mathrm{EUV}$ profile the $X_\mathrm{radio}$ interval from 1.4 to 1.5 with black crosses (Figure~\ref{PrevProfileDirect}).

\subsection{Temporal Evolution of the EUV Compression Ratio}

We apply the method described in the previous section to several 211\,\AA\,images during the bubble expansion (from 05:38 to 05:40 UT) for different angles. Our aim is to search for connections between  $X_\mathrm{radio}$ and $ X_\mathrm{EUV}$, which in turn will allow us to draw connections between the metric Type-II shock and the EUV wave.

\begin{figure}\centering
\centerline{ \includegraphics[width=0.95 \textwidth]{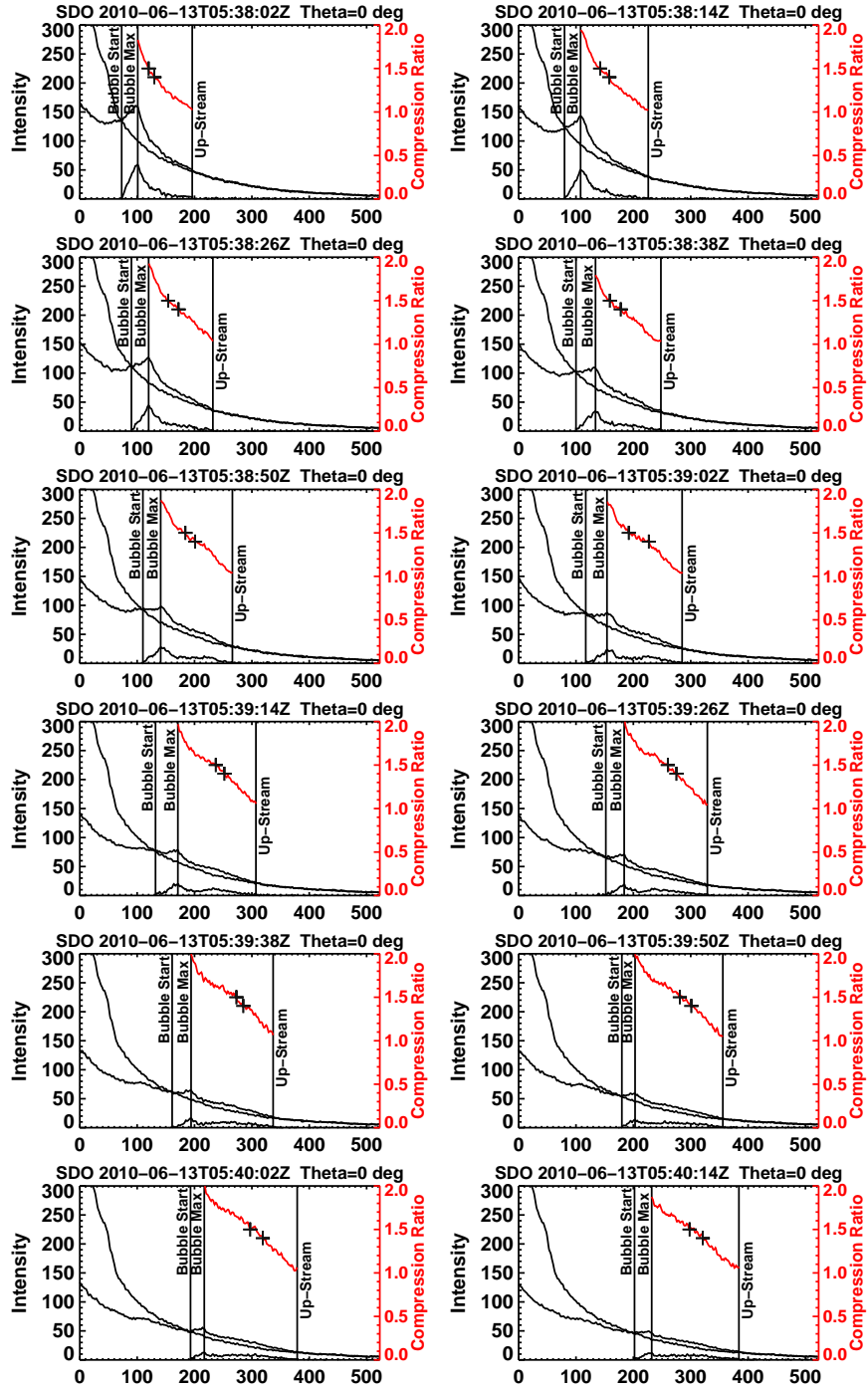}}
   \caption{Measurements of the EUV intensity profiles and $ X_\mathrm{EUV}$ along the $\theta=0^\circ$ direction for 211~\AA, from 05:38:02 to 05:40:14 UT. The vertical lines from left to right correspond to the start of the bubble, the bubble maximum and the inner boundary of the sheath region (see also Figure~\ref{PrevProfileDirect} and the discussion of Section~\ref{EUVcompDisc}). The calculated EUV compression ratio [\,$X_\mathrm{{EUV}}$\,] for the region between the bubble maximum and the wave upstream region is displayed with the red line. The black crosses mark the measured interval of $X_\mathrm{{radio}}$ values during the observed Type II. The $x$-axes is in pixels along the corresponding path (see Figure~\ref{PrevImgCompRatioDirect}).}
         \label{CompPrevThetO}
   \end{figure}

In Figure~\ref{CompPrevThetO} we present the evolution of the EUV intensity profiles and the compression ratio for the direction of the bubble radial expansion. From the EUV intensity profiles and the difference curve between the background and the event intensity profile, we examined the evolution of the bubble and the wave formation and propagation. The intensity of the bubble continuously weakens from 05:38:02 to 05:39:38 when it reaches an almost constant intensity. The wave starts to form at 05:37\todash{}05:38 UT and it is best observed in the difference curve after 05:38:38 UT. Its intensity continuously increases probably gaining energy from its driver, until 05:39:26. After that time, its intensity slightly drops until it reaches an almost constant value. This drop of the intensity is accompanied by a broadening of the wave region which possibly means that around this time the wave probably detaches from its driver and the corresponding shock wave may not be driven any more.

The values of compression ratio ($ X_\mathrm{EUV}$ in Figure~\ref{CompPrevThetO}) as computed from the analysis of the previous section, vary between one and two in the region between the bubble maximum and wave end. The lower value of the $ X_\mathrm{EUV}$ (=1) simply reflects the compression ratio at the inner boundary of the upstream region. The upstream region and its intensity are used for the computation of the $ X_\mathrm{EUV}$ profile.

We compare the values of the $ X_\mathrm{EUV}$ profile with the $X_\mathrm{radio}$ (crosses in Figure~\ref{CompPrevThetO}). When the wave starts to form at 05:38 UT the $X_\mathrm{radio}$ values are mostly localised within the region of the bubble end and the the wave bump. This localization can be observed until 05:39:02 UT.  After this time the $X_\mathrm{radio}$ slightly starts to drift well after the bubble end and mostly localised at the region of the wave bump until the end of this analysis at 05:40:14 where the wave have reach the field of view (FOV) of 211\,\AA\, AIA images.

\begin{figure}\centering
\centerline{ \includegraphics[width=0.95 \textwidth]{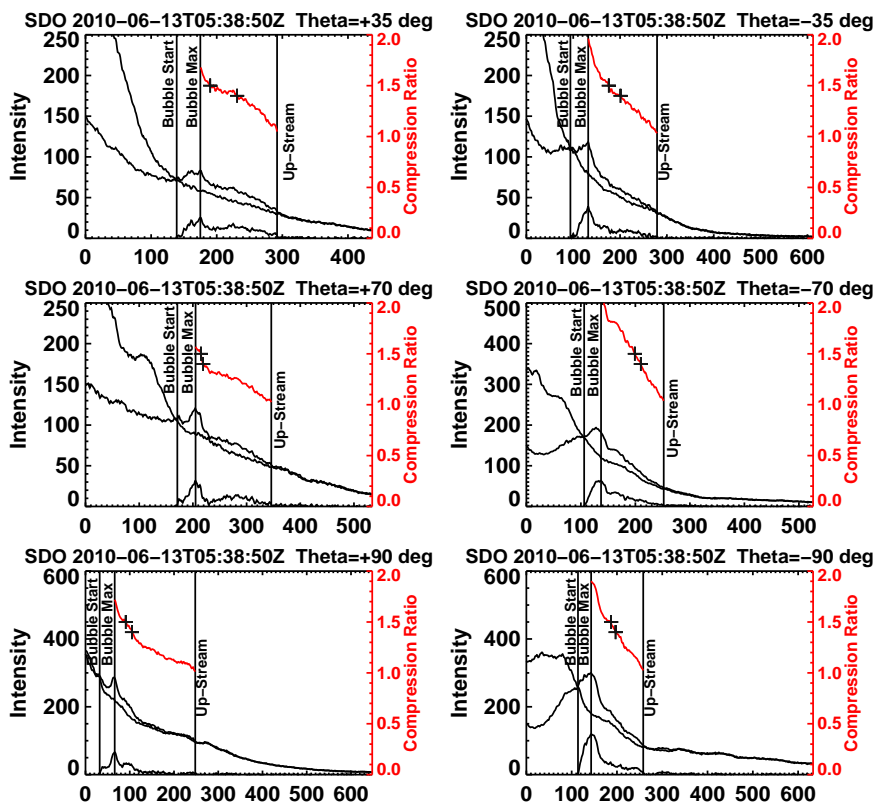}}
   \caption{Measurements of the EUV intensity profiles and $ X_\mathrm{EUV}$ at different angles from $\theta=0^\circ$ direction (\ie~ $\pm 35^\circ \pm 70^\circ \pm 90^\circ$, see Figure~\ref{PrevImgCompRatioDirect}), for 211~\AA, at 05:38:50 UT. The vertical lines from left to right correspond to the start of the bubble, the bubble maximum and the inner boundary of the sheath region (see also Figure~\ref{PrevProfileDirect} and the discussion of Section~\ref{EUVcompDisc}). The calculated EUV compression ratio [\,$X_\mathrm{{EUV}}$\,] for the region between the bubble maximum and the wave upstream region is displayed with the red line. The black crosses mark the measured interval of $X_\mathrm{{radio}}$ values during the observed Type II. The $x$-axes is in pixels along the corresponding path (see Figure~\ref{PrevImgCompRatioDirect}).}
         \label{ComprtoAngle}
   \end{figure}

We repeated the above analysis for several directions away from the radial to investigate the EUV compression ratio in lateral directions with respect to the bubble expansion (Figure~\ref{ComprtoAngle}). We ordained similar results as for the radial direction for the temporal profiles of the event EUV intensities and compression ratio. Significant differences were only observed for directions perpendicular to the radial direction where deflected background structures blend with the bubble and EUV wave profile. For these directions the determination of the bubble start\todash{}maximum and wave upstream region becomes more uncertain. A sample of the resulting EUV profiles and compression ratio is presented in Figure~\ref{ComprtoAngle} for a selected time for all the considered directions.

To sum up, the above results indicate that the bubble gets fainter with distance and the EUV wave clearly seems to detach from its driver (bubble) before it exits the AIA FOV. The sheath region (\ie~between the bubble max and the inner boundary of the upstream region) becomes broader during the expansion. For all frames $X_\mathrm{radio}$ (black crosses, Figure~\ref{CompPrevThetO}) fall within the sheath region (red line, Figure~\ref{CompPrevThetO}) deduced from the $ X_\mathrm{EUV}$. Our results also suggest that the computed $ X_\mathrm{EUV}$ is consistent with the measured $X_\mathrm{radio}$ in sheath regions both along the radial and the lateral direction (Figure~\ref{ComprtoAngle}).

The EUV wave/shock decelerating character, the decrease of its amplitude and increase of its width with time (pulse spread), and the increasing bubble-shock stand-off distance all point to a freely propagating wave. However, for a period early in the event, when the stand-off distance between the bubble-shock is small, and it is hard to distinguish between the blast-wave and the piston driven nature.

\section{Discussion and Conclusions}\label{Disc}
We have combined observations of SDO/AIA images and high-resolution dynamic spectra obtained by the ARTEMIS IV radio spectrograph of the 13 June 2010 eruptive event. We performed a joint analysis of high-cadence EUV imaging and radio spectral observations to infer the nature of the shock driver responsible for the observed Type-II burst. To connect the evolutions and structures observed with the AIA imaging with the Type-II burst we performed the following:

\begin{itemize}
\item We introduce a new method to calculate a compression-ratio proxy from the EUV images of AIA. From the comparison of the $X_\mathrm{radio}$ and  $X_\mathrm{EUV}$ we found that the Type-II radio burst could originate in the sheath region between the bubble (wave driver) and the EUV shock front, in both radial and lateral directions.

\item From the comparison between the height--time measurements of the bubble and the EUV wave, both calculated along the radial direction, and the Type-II height from the frequency drift we observe that the Type-II heights fall within the sheath region between the bubble (EUV wave driver) and the EUV shock front in the radial direction.

\item The Type-II radio burst starts around the lateral over-expansion phase of the bubble, which suggests that this phenomenon could play an important role into driving the shock.

\end{itemize}

Our results give further support to the role of the lateral over-expansion of CMEs in driving wave and shock phenomena that are observed in various spectral domains in the inner corona.

\begin{acks_M}
{The authors would like to thank, Victor Grechnev for useful discussion and the anonymous referee for his comments, that helped to improve the manuscript. This research was supported in part by the European Union (European Social Fund-ESF) and Greek national funds through the Operational Program ``Education and Lifelong Learning" of the National Strategic Reference Framework (NSRF) - Research Funding Program: Thales. Investing in knowledge society through the European Social Fund. SP acknowledges support from an FP7 Marie Curie Re-integration Grant (FP7-PEOPLE-2010-RG/268288). AV's work is supported by NASA and ONR funds. SDO is a mission of NASAs Living With a Star Program. The SDO data are provided as a courtesy of NASA/SDO and the AIA science teams. SOHO is a project of international cooperation between ESA and NASA. The LASCO CME catalog is generated and maintained at the CDAW Data Center by NASA and The Catholic University of America in cooperation with the Naval Research Laboratory.} 
\end{acks_M}


\end {article}
\end {document}